%% file: manuscript.tex
\documentclass[pdflatex,sn-nature]{sn-jnl}% Style for submissions to Nature Portfolio journals
\usepackage{geometry}
\usepackage{graphicx}%
\usepackage{multirow}%
\usepackage{amsmath,amssymb,amsfonts}%
\usepackage{amsthm}%
\usepackage{mathrsfs}%
\usepackage[title]{appendix}%
\usepackage{xcolor}%
\usepackage{textcomp}%
\usepackage{manyfoot}%
\usepackage{booktabs}%
\usepackage{algorithm}%
\usepackage{algorithmicx}%
\usepackage{algpseudocode}%
\usepackage{listings}%
\input{packages_and_macros}

\hypersetup{bookmarksdepth=2}

\begin{document}
\title[]{DeepPD: Joint Phase and Object Estimation from Phase Diversity with Neural Calibration of a Deformable Mirror}

% object estimation / aberration correction


\author[1]{\fnm{Magdalena C.} \sur{Schneider}}\email{schneiderm2@janelia.hhmi.org}

\author[1]{\fnm{Courtney} \sur{Johnson}}
\author[1]{\fnm{Cedric} \sur{Allier}}
\author[1]{\fnm{Larissa} \sur{Heinrich}}
\author[1]{\fnm{Diane} \sur{Adjavon}}
\author[2]{\fnm{Joren} \sur{Husic}}
\author[2]{\fnm{Patrick} \sur{La Rivi\`{e}re}}
\author[1]{\fnm{Stephan} \sur{Saalfeld}}
\author[1]{\fnm{Hari} \sur{Shroff}}
% \author[1]{\fnm{et} \sur{al.?}}

% \author[1]{\fnm{et} \sur{al.}}%\email{iiauthor@gmail.com}


% Department: \orgdiv{}
\affil[1]{\orgname{HHMI Janelia Research Campus}} % , \orgaddress{\street{Street}, \city{City}, \postcode{100190}, \state{State}, \country{Country}}}

\affil[2]{\orgname{University of Chicago}}
%\orgaddress{\street{Street}, \city{City}, \postcode{10587}, \state{State}, \country{Country}}}

%\affil[3]{\orgdiv{Department}, \orgname{Organization}, \orgaddress{\street{Street}, \city{City}, \postcode{610101}, \state{State}, \country{Country}}}


\abstract{\input{sections/00_abstract}}




\maketitle

\input{sections/01_introduction}

\section{Results}\label{sec2}
\input{sections/02_results}

\section{Discussion}\label{sec12}
\input{sections/03_discussion}

\backmatter

%\bmhead{Supplementary information}
%Info on supplements

%\bmhead{Acknowledgements}

% \clearpage
\section{Methods}\label{sec:methods}
\input{sections/04_methods}

%%%%%%%%%%%%%%%
\section*{Declarations}
%Some journals require declarations to be submitted in a standardised format. Please check the Instructions for Authors of the journal to which you are submitting to see if you need to complete this section. If yes, your manuscript must contain the following sections under the heading `Declarations':

\subsection*{Competing interests}
M.C.S, C.J., C.A., P.J.L., and H.S.\ filed a patent application for the technique described here.

\subsection*{Code availability}
Code for phase diversity analysis using the Gaussian method is available at \url{https://github.com/ceej640/PhaseDiversity}.
Code for phase diversity analysis (Poisson and neural representation methods) and mirror model training
will be made freely available via GitHub upon publication of the manuscript.

\subsection*{Author contribution}
M.C.S.\ and H.S.\ supervised the project, with S.S.\ providing general oversight and contributing to discussions.
C.J.\ and H.S.\ designed and built the microscope setup.
C.J.\ performed all microscopy data acquisition.
M.C.S., J.H., and P.J.L.\ implemented the Poisson algorithm.
M.C.S.\ and C.A.\ conceived the idea of a mirror model.
M.C.S.\ designed the network architecture of the mirror model with input from L.H.\ and D.A., designed the loss function and training paradigm, and performed training.
M.C.S.\ and C.A.\ implemented an initial version of the code for phase diversity analysis using neural representations and the mirror model and conducted initial experiments using neural representations.
M.C.S.\ performed further development and conducted all final computational experiments.
M.C.S.\ wrote the manuscript with input from all authors.

\subsection*{Acknowledgments}
This work was supported by the Howard Hughes Medical Institute. This article is subject to HHMI’s Open Access to Publications policy. %which grants a nonexclusive CC BY 4.0 license to the public and a sublicensable license to HHMI in the research articles.
We thank Christopher A.\ Metzler and Ashok Veeraraghavan for valuable discussions on the NeuWS method.

%\begin{itemize}
%\item Conflict of interest/Competing interests %(check journal-specific guidelines for which heading to use)
%\item Ethics approval and consent to participate
%\item Consent for publication
%\item Data availability 
%\end{itemize}

%\noindent
%If any of the sections are not relevant to your manuscript, please include the heading and write `Not applicable' for that section. 


\clearpage
\bibliography{references}

\clearpage
% Set counters for SI
\pagenumbering{gobble}
\setcounter{page}{1}
\renewcommand{\thepage}{S\arabic{page}}
\setcounter{section}{0}
\setcounter{figure}{0}
\input{sections/SI}

\end{document}

%% file: packages_and_macros.tex
\usepackage{siunitx}
\usepackage{amsmath}

\usepackage{nameref}

\newcommand{\NA}{\textrm{NA}}
\newcommand{\pixelsize}{s_{\textrm{px}}}
\newcommand{\loss}[1]{\mathcal{L}_{\textrm{\footnotesize #1}}}
\newcommand{\panel}[1]{\textbf{(#1)}}
\newcommand{\subpanel}[1]{#1}
\newcommand{\fourier}[1]{\mathcal{F}(#1)}
\newcommand{\fourierinv}[2]{\mathcal{F}^{-1}#2( #1 #2)}
\newcommand{\objest}{\hat{O}}
\newcommand{\obj}{O}
\newcommand{\phaseest}{\hat{\phi}}
\newcommand{\phase}{\phi}
\newcommand{\phasediv}{\psi}
\newcommand{\PSF}{\textrm{PSF}}
\newcommand{\OTF}{\textrm{OTF}}

\newcommand{\Lsupervision}{\mathcal{{L}_{\text{sup}}}}
\newcommand{\Lcycle}{\mathcal{{L}_{\text{cc}}}}

\newcommand{\Lregularization}{\mathcal{{L}_{\text{reg}}}}

\newcommand{\Lmirror}{\mathcal{{L}_{\text{mirror}}}}

\newcommand{\RMSWD}{\textrm{RMS}_{\text{W}}}

\newcommand{\SSIM}{\textrm{SSIM}}

\newcommand{\MSE}{\textrm{MSE}}
\newcommand{\DCT}{\textrm{DCT}}
\newcommand{\dct}{\mathcal{F}_c}
\newcommand{\abslog}[1]{\textrm{abslog}_{#1}}

\makeatletter
\renewcommand{\email}[1]{%
  \global\advance\emailcnt by 1\relax
  \if@corauemail
     \g@addto@macro\corrauthemail{%
        \setcounter{footnote}{0}%
        \textcolor{blue}{#1}%
     }%
  \else
     \g@addto@macro\authemail{%
        \setcounter{footnote}{0}%
        \textcolor{blue}{#1}%
     }%
  \fi}
\makeatother

%% file: sections/00_abstract.tex
%%% Abstract

Sample-induced aberrations and optical imperfections limit the resolution of fluorescence microscopy.
Phase diversity is a powerful technique that leverages complementary phase information in sequentially acquired images with deliberately introduced aberrations---the phase diversities---to enable phase and object reconstruction and restore diffraction-limited resolution.
These phase diversities are typically introduced into the optical path via a deformable mirror.
Existing phase-diversity-based methods are limited to Zernike modes, require large numbers of diversity images, or depend on accurate mirror calibration---which are all suboptimal.
We present DeepPD, a deep learning–based framework that combines neural representations of the object and wavefront with a learned model of the deformable mirror to jointly estimate both object and phase from only five images. DeepPD improves robustness and reconstruction quality over previous approaches, even under severe aberrations. We demonstrate its performance on calibration targets and biological samples, including immunolabeled myosin in fixed PtK2 cells.

%% file: sections/01_introduction.tex
%%% Introduction

\section{Introduction}\label{sec:intro}

The potential of light microscopy is often unrealized due to aberrations introduced either by imperfections in the optical system or by the sample itself, yielding blurry and distorted images that fail to achieve diffraction-limited resolution.
The goal of adaptive optics is to estimate the aberrations present in an image and counter them during acquisition via a deformable mirror (DM) or spatial light modulator \cite{Hampson2021}, or, in computational settings, to estimate and correct them post-acquisition.
Aberrations are caused by variations in the phase of the optical field. Yet the phase cannot be measured directly, as conventional imaging systems only capture light intensity.
Estimating the phase is a challenging problem, and various wavefront estimation methods have been proposed to address this issue. These methods can be broadly classified into two categories: guidestar-based and guidestar-free methods \cite{Hampson2021}.

Guidestar-based methods rely on the presence of a point source in the sample, and the phase aberration can be retrieved, for example, by a Shack-Hartmann (SH) wavefront sensor \cite{Hampson2021} that measures the phase aberration based on displacements of the point source's images on a microlens array.
However, a point source is not always readily available in a sample, and introducing one can be difficult, expensive or obstruct the observation of the actual sample of interest.
While image-based SH methods that do not require a point source exist \cite{Poyneer2003}, they are less robust.
Furthermore, the SH sensor adds to the cost of the setup, and needs to be placed in a separate light path, which can introduce additional aberrations or misalignment errors.

Guidestar-free methods reconstruct the phase aberration directly from one or multiple images of the object and do not require a point source.
While a variety of deep learning methods have been proposed to estimate the aberration based on a single image \cite{Kang2024,Guo2025}, the problem of joint sample and phase estimation is inherently ill-posed, and the performance of the neural networks depends on a strong prior on the sample space, which can limit generalization to unseen samples.
To alleviate this ill-posedness, other methods turn to recording multiple images per sample, with each image containing different information---either by changing the illumination pattern \cite{Wijethilake2023} or by introducing phase diversity \cite{Feng2023}.
While this alleviates the need for a strong prior on the sample space, it comes at the cost of increased acquisition time for 32 and 100 images for \cite{Wijethilake2023} and \cite{Feng2023}, respectively, and the potential for increased phototoxicity.
To address these limitations, Johnson et al.\ recently introduced a guidestar-free method that achieves high-quality phase retrieval and aberration correction from only five images, including the aberrated image and four phase-diversity images \cite{Johnson2024a}.
While this method can correct a wide range of aberrations, it relies on an analytic calculation of the gradient and Hessian for the Gauss-Newton optimization algorithm based on a limited set of Zernike modes, which can limit performance for higher-order aberrations \cite{Liu2025}.

More broadly, the effectiveness of phase-diversity-based approaches hinges on the controlled introduction of deliberate aberrations---the phase diversities---added to the unknown aberration.
These can be introduced, for example, via a DM as in \cite{Johnson2024a}.
Precise control over the introduced aberrations is critical for the success of phase-diversity-based methods as a mismatch between the intended and actual aberrations can distort the solution space and prevent the optimization algorithm from converging to the correct solution---a challenge that, to our knowledge, has not been addressed previously.
While the response of DMs is typically assumed to be linear, this is not necessarily true in practice \cite{Haber2021}, leading to suboptimal performance.

The correction of aberrations is typically performed with a DM in the emission light path.
Imperfect calibration of the DM will lead to suboptimal correction results, even in the case of perfect phase retrieval.
Notably, reconstruction and correction would ideally be performed in real-time.
For rapidly moving samples, this can be challenging—especially with large images, which are computationally more expensive, and high aberration magnitudes, which increase the time required for the analysis to converge.
Instead of being used solely for phase estimation, the phase-diversity images can also be used to estimate the object itself. This estimation can be performed after acquisition, when the time budget for reconstruction is not tightly constrained.

Here, we present DeepPD, a combined approach that integrates neural representations \cite{Feng2023}, selected phase diversities \cite{Johnson2024a}, and a learned deformable mirror calibration model. This enables high-order aberration estimation while reducing acquisition time.
The phase and object estimates are jointly optimized by minimizing the difference between predicted and acquired images.
We implemented the imaging system as a fully differentiable model in PyTorch, enabling gradient-based optimization via automatic differentiation.
To address the limitations of linear mirror calibration, we introduce a novel approach for DM calibration that employs deep neural networks to model the mirror response, enhancing the model to capture higher-order aberrations and nonlinear effects.
DeepPD accurately reconstructs both the phase aberration and the object from only five images. We demonstrate its performance on a calibration slide with known object shapes and on biological samples of myosin-labeled PtK2 cells.

%% file: sections/02_results.tex
%% Results

\begin{figure}[!htb]
    \centering
    \includegraphics[width=0.9\textwidth]{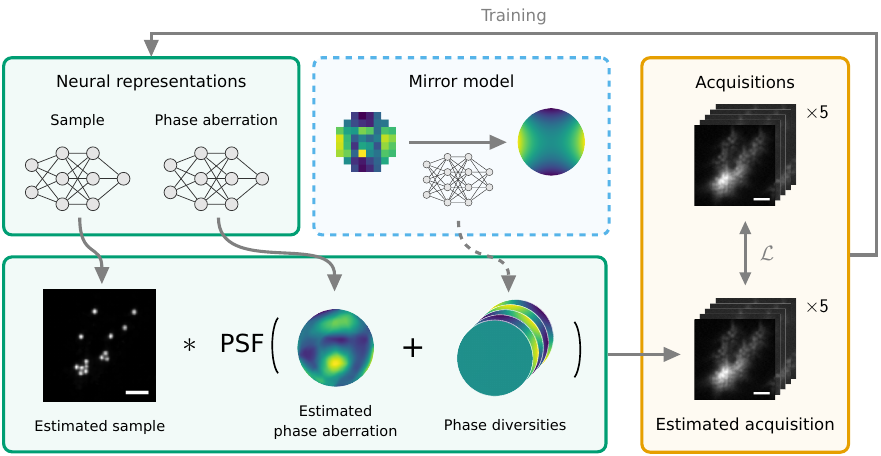}
    \vspace{5pt}
    \caption{DeepPD method overview. The schematic depicts the concept of estimating the unknown object and phase aberration based on phase diversity.
    The estimated sample and estimated phase aberrations are represented by neural representations.
    The phase diversities are obtained from the input voltages based on a pretrained mirror model.
    By convolving the current object estimate with PSFs based on the current phase aberration estimate and the input phase diversities we obtain an estimate of the aberrated image and the four phase-diversity images (i.e., a total of five images).
    Note that the aberrated image only includes the estimated phase aberration and no additional phase diversity (indicated by the flat phase).
    The obtained estimated acquisition is compared to the acquired images to calculate the loss, which is used to train the neural representations. Scale bars: $\SI{1}{\um}$.
    }
    \label{fig:method-overview}
\end{figure}

DeepPD jointly estimates the phase aberration and object from five images.
The set of acquired images for our approach consists of an aberrated image (deteriorated by an unknown phase aberration) and four additional phase-diversity images.
To obtain the phase-diversity images, we induce additional known phase modulations, adding them to the unknown aberration via a deformable mirror.
For these induced phase modulations we selected a set of four astigmatism diversities (oblique and vertical astigmatism with opposite signs) that were previously shown to yield the best performance among the tested diversities \cite{Johnson2024a}.
The reconstruction procedure for retrieving the object and phase aberration estimate is depicted in Fig.~\ref{fig:method-overview} and is based on the imaging equation
\begin{equation}
    I_k = \obj * |\mathcal{F}^{-1}(P\,e^{i(\phase+\phasediv_k)})|^2, \quad \text{for } k \in \{0,\dots,4\},
    \label{eq:imaging-equation}
\end{equation}
where $I_k$ is the acquired image, $\obj$ the object, $P$ the pupil aperture, $\phase$ the (unknown) phase aberration, $\phasediv_k$ the $k$th phase diversity (with $\psi_0 \equiv 0$), and $*$ denotes convolution.

Both the object $\obj$ and the phase aberration $\phase$ are unknown and need to be retrieved.
Instead of representing the object and phase aberration as arrays and optimizing each pixel individually, we use neural representations as introduced by NeuWS \cite{Feng2023}.
The neural representations are neural networks that take spatial or frequency coordinates as input, and output the object or phase aberration, respectively.
whereas NeuWS requires around $100$ images with random aberrations as phase diversities, we use four selected diversities, which considerably reduces acquisition time.
Moreover, we square the output of the last layer of the object neural representation to prevent nonphysical negative values in the object estimate and to improve training.
The details of the neural network architectures are described in the \nameref{sec:methods}.
Neural representations can be evaluated at any continuous spatial or frequency coordinate, and are not limited to the discrete pixel grid of the acquired images.
The neural networks representing the object and phase estimate are trained to minimize the discrepancy between the estimated images---based on the current object and phase estimates---and the acquired images.

Since the phase diversities are introduced via the deformable mirror, any inaccuracies in the mirror calibration will directly affect and alter the set of estimated images.
The solution space for the object and phase aberration will therefore be compromised, preventing the training of the neural representations from converging to the correct solution.
Instead of using the assumed induced astigmatism, we estimated the applied wavefront $\phasediv_k$ for each diversity via a pretrained mirror model that predicts the wavefront from the voltages applied to the mirror actuators.

\begin{figure}[!htb]
    \centering
    \includegraphics[width=\textwidth]{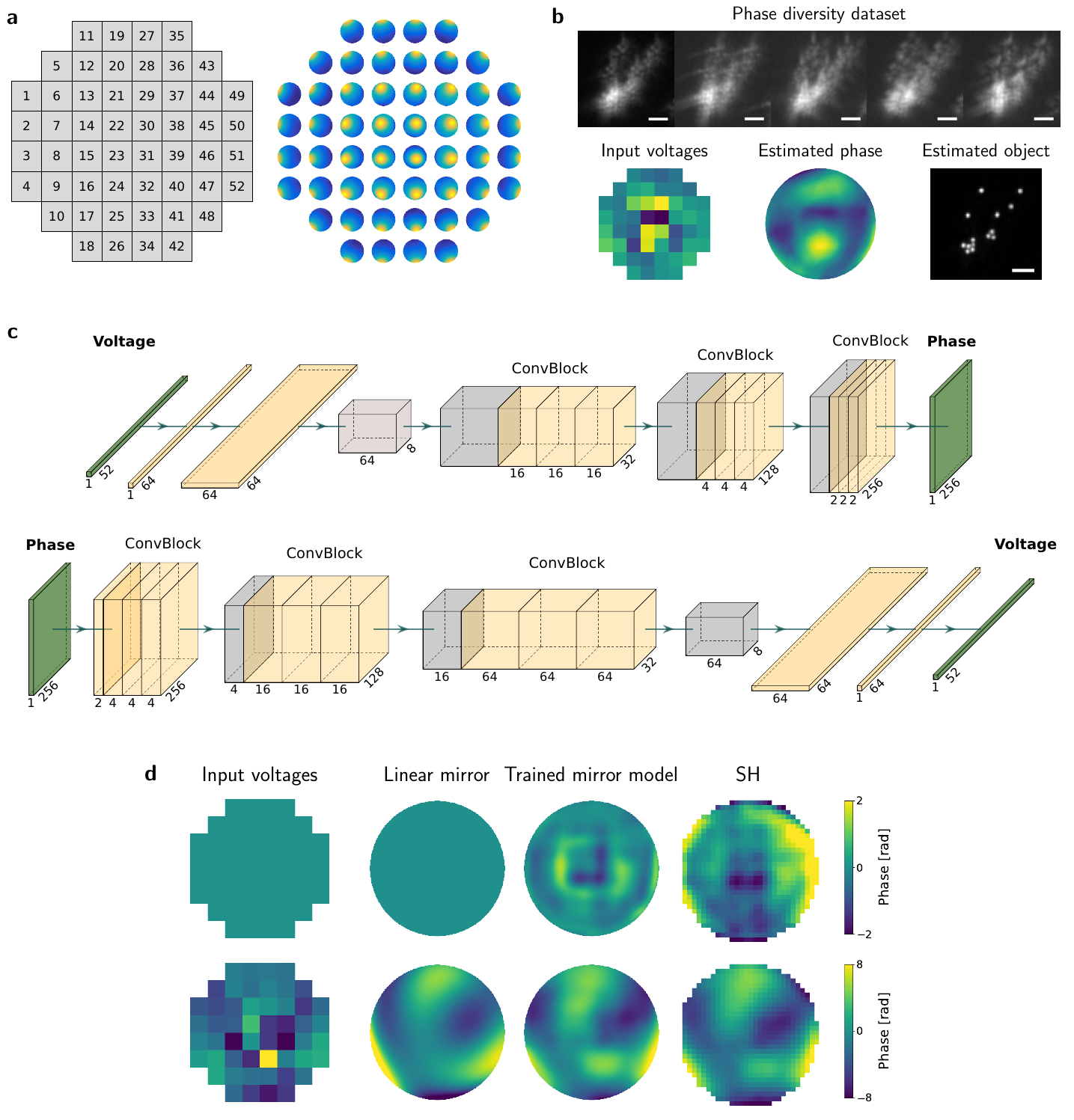}
    \vspace{-5pt}
    \caption{Learned mirror calibration. \panel{a} Applying voltage to each of the 52 actuators (left) shapes the mirror's deformable membrane. Numbers indicate the mirror actuator indices. Assuming linearity, the mirror's response to increasing the voltage for a single actuator can be described by an influence function (right).
    \panel{b} Phase-diversity images of a training data set. Data is recorded on a bead sample.
    The phase aberration is estimated from a phase-diversity acquisition data set consisting of the aberrated image and four phase-diversity images (top).
    The input voltages (bottom left) together with the obtained phase estimates (bottom center) from the neural representations (Fig.~\ref{fig:method-overview}) constitute one training data pair for the mirror model. The object estimate is not used for training the mirror model, but is shown here for illustration (bottom right).
    Scale bars: $\SI{2}{\um}$.
    \panel{c} Network architecture of the mirror model for predicting (non-linear) phase response from voltage input (top) and obtaining required input voltages for a desired phase response (bottom). Details of the architectures are described in the \nameref{sec:methods}.
    \panel{d} Wavefront induced by the deformable mirror for two sets of applied voltages.
    The left column shows the input voltages applied to the 52 mirror actuators. Columns 2--4 compare the expected wavefront by linear combination of the influence functions, predicted mirror response by the trained neural-network mirror model, and SH measurements of the phase aberration. Note that the SH measurements are shown in a zonal representation, and the pupil diameter cannot be matched directly.
    }
    \label{fig:mirror-model}
\end{figure}

\paragraph{Mirror model.}
The response of a DM is typically assumed to be a linear combination of a set of influence functions that describe the mirror's response to each actuator (Fig.~\ref{fig:mirror-model}\subpanel{a}).
Those influence functions are usually either expressed via a Zernike basis or mirror modes \cite{Wang2009}.
However, nonlinear deformations---especially over wider voltage ranges---and couplings between the actuators lead to non-linearities in the mirror response \cite{Haber2021} that cannot be accurately described by a linear model.

To address this issue, we designed a neural network as part of DeepPD that predicts the phase response of our MIRAO 52ES (Imagine Optic) deformable mirror based on the voltages applied to the 52 mirror actuators. 
A second network with a reversed architecture allows to retrieve the required voltages for a desired phase response.
We refer to the two parts of the mirror model as voltage-to-phase and phase-to-voltage networks, respectively.
Both network architectures are depicted in Fig.~\ref{fig:mirror-model}\subpanel{c} and described in detail in Section~\ref{sec:methods-nn-architecture} of the \nameref{sec:methods}.

Training of the mirror model requires a set of training data pairs consisting of the applied voltages and the corresponding phase response.
As no direct measurement of the phase response is available, we estimated the phase response from sets of phase-diversity images acquired on a fluorescent bead sample (Fig.~\ref{fig:mirror-model}\subpanel{b}).
For this purpose, we recorded $\num{3930}$ sets of phase-diversity images, each with a random phase aberration introduced by the deformable mirror (see \nameref{sec:methods} for details).
In order to induce the phase diversities, we relied on an initial linear mirror calibration as described in \cite{Johnson2024a}.
We analyzed these datasets with our method based on neural representations to obtain the phase estimates. 
As the mismatch between the assumed and the actual phase diversities due to the imperfect mirror calibration may introduce errors in the obtained phase estimates, we trained the mirror model in three cycles of estimating the phase response and (re-)training the mirror model, after which no further improvement was observed.
The phase diversities used for analysis in the second and third cycle were obtained by evaluating the mirror model of the previous cycle.
To ease the estimation of the correct phase response, we recorded a separate image without any aberration for initializing the neural representation of the object.
The estimated phase response was initialized with the phase estimate obtained by the linear mirror calibration in the first cycle, and with the phase estimate obtained by the previously trained mirror model in the subsequent cycles.

Additional imperfections in the phase estimates may arise from the phase-diversity method itself.
For example, tip or tilt induced by the mirror cannot be distinguished from a lateral shift of the object in the sample plane.
Noise in the phase estimates causes overfitting and prevents the mirror model from generalizing well to unseen data when training the voltage-to-phase network. 
To compensate for these issues and to improve generalization, we trained both the voltage-to-phase and phase-to-voltage networks together and included a cycle-consistency loss:
passing a voltage through the voltage-to-phase network and then through the phase-to-voltage network should yield the original voltage;
passing a phase through the phase-to-voltage network and then through the voltage-to-phase network should yield the original phase.

\paragraph{Mirror model matches SH wavefronts better than linear mirror model.}
To evaluate the performance of our trained mirror model, we compared the predicted phase aberration to measurements from a SH sensor (HASO4, Imagine Optics) on single beads.
The SH sensor provides a direct measurement of the wavefront, and is considered the gold-standard for wavefront sensing.
However, the resolution of the SH measurement is limited by the number of lenslets in the sensor.
Moreover, the sensor needs to be placed in a separate light path that may introduce additional aberrations not present in the imaging path.
Therefore, a quantitative comparison provides limited value and we resorted to a qualitative comparison instead.

Fig.~\ref{fig:mirror-model}\subpanel{d} compares the predicted wavefront from both a linear calibration and the trained mirror model to the SH measurements.
For no applied voltage, the linear mirror model \cite{Johnson2024a} predicts a flat phase, while the trained mirror model predicts a phase with $\SI{44}{\nm}$ root-mean-square (RMS) wavefront error and a pattern closely matching the SH measurements (Fig.~\ref{fig:mirror-model}\subpanel{d}).
Also for non-zero voltages, the trained mirror model predicts a phase that is more similar to the SH measurements than the linear mirror model.

\paragraph{Neural representations and a trained mirror network improve phase retrieval and object estimation on Argo-HM calibration slide.}
Next, we investigated the performance of our approach compared to previous methods \cite{Johnson2024a,Reiser2023} for phase retrieval and object estimation.
Since ground truth for the object is typically unavailable for biological samples, we initially validated our approach using an Argo-HM calibration slide (Argolight), which is a manufactured glass slide with known fluorescent patterns embedded inside.
We compared our method against three other approaches:
(i) a Gauss--Newton algorithm assuming Gaussian noise based on analytic derivatives of the PSF and a Zernike representation of the phase \cite{Johnson2024a},
(ii) an iterative optimization algorithm assuming Poisson noise statistics and a Zernike representation of the phase \cite{Reiser2023},
and (iii) iterative optimization based on automatic differentiation using neural representations of the object and phase, and assuming a linear mirror response (details on implementations in \nameref{sec:methods}). For convenience, we henceforth refer to these methods as Gaussian, Poisson, NNRs, and our full method as DeepPD.

\begin{figure}
    \centering
    \includegraphics[width=\textwidth]{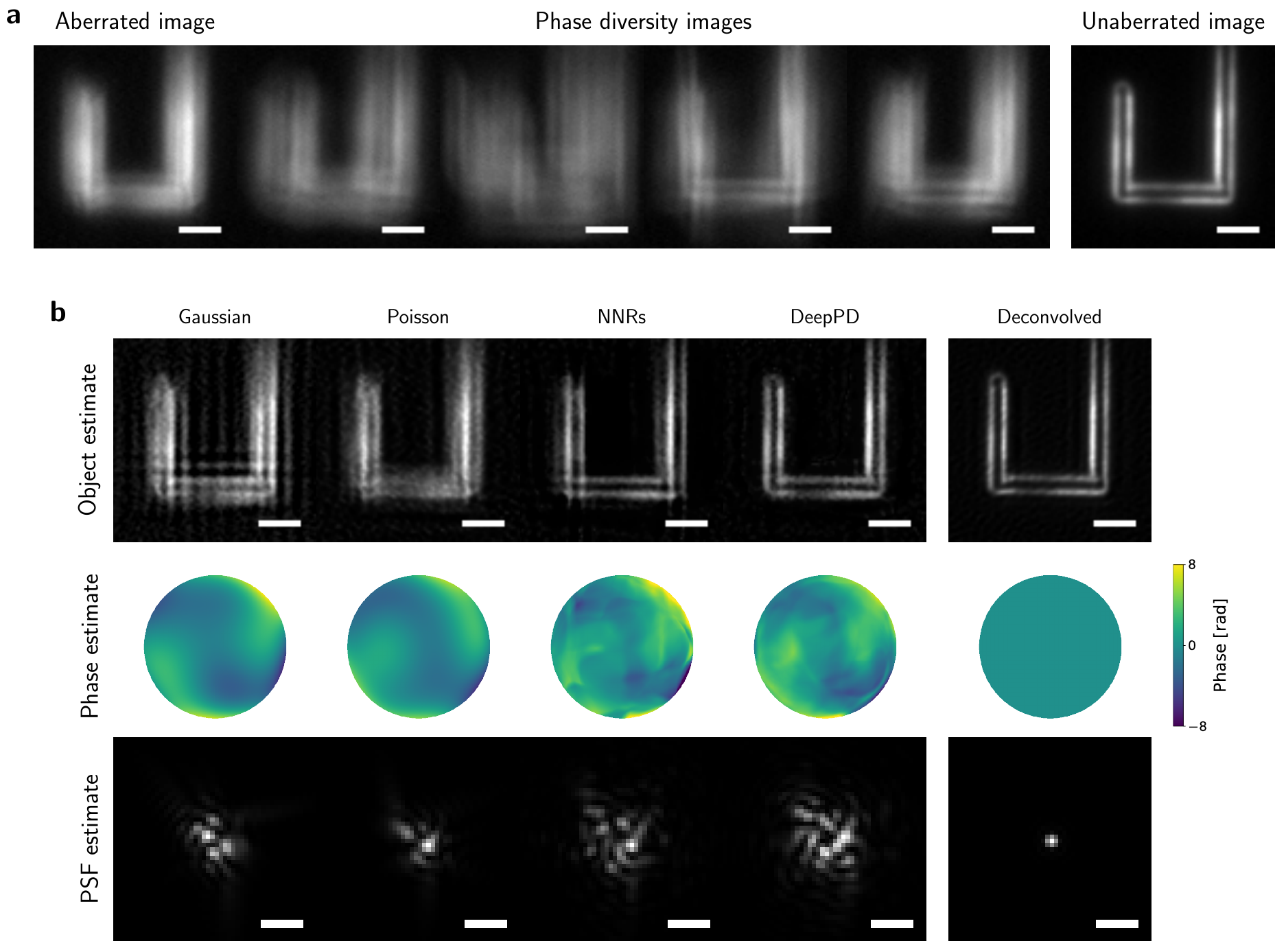}
    \caption{Performance comparison of various methods on the Argo-HM calibration slide.
    \panel{a} Recorded microscope images. Left to right: Aberrated image, four phase-diversity images, and the unaberrated image for comparison. Scale bars: $\SI{2}{\um}$.
    \panel{b} Object estimate (top row), phase estimate (middle row) and corresponding PSF (bottom row) obtained by various methods (left to right: Gauss--Newton algorithm, Poisson algorithm, neural representations assuming a linear mirror response (NNRs), and neural representations using the trained mirror model (DeepPD)). The last column shows an object estimate obtained by Richardson--Lucy deconvolution (20 iterations) of the unaberrated image, flat phase and corresponding PSF for comparison. Scale bars: $\SI{2}{\um}$ (object estimates), $\SI{1}{\um}$ (PSFs).
    }
    \label{fig:argolight}
\end{figure}

Fig.~\ref{fig:argolight}\subpanel{a} shows the recorded aberrated image, and the corresponding four phase-diversity images obtained with astigmatism diversities for one region of interest.
For comparison, we also show the unaberrated image. The object is part of the letter G in the word ARGOLIGHT on the calibration slide. The letter is outlined by a double line, separated by a distance of $\SI{600}{\nm}$.
The aberrated image is obtained by inducing a random aberration via the deformable mirror which is assumed to be unknown for purpose of the analysis.
Fig.~\ref{fig:argolight}\subpanel{b} shows a comparison of the various object and phase estimates by the Gaussian, Poisson, NNRs, and DeepPD.
While the objects retrieved by the first three methods show artifacts, DeepPD reconstructs the correct shape of the object.
For comparison, we also show an object estimate obtained by Richardson--Lucy deconvolution of the unaberrated image, which closely resembles the object estimate obtained by our method.
As both NNRs and DeepPD are based on neural representations of the phase, they are not restricted to the first $21$ Zernike modes for estimating the phase and are thus able to capture finer details of the PSF than the Gaussian and Poisson methods.

\begin{figure}
    \centering
    \includegraphics[width=\textwidth]{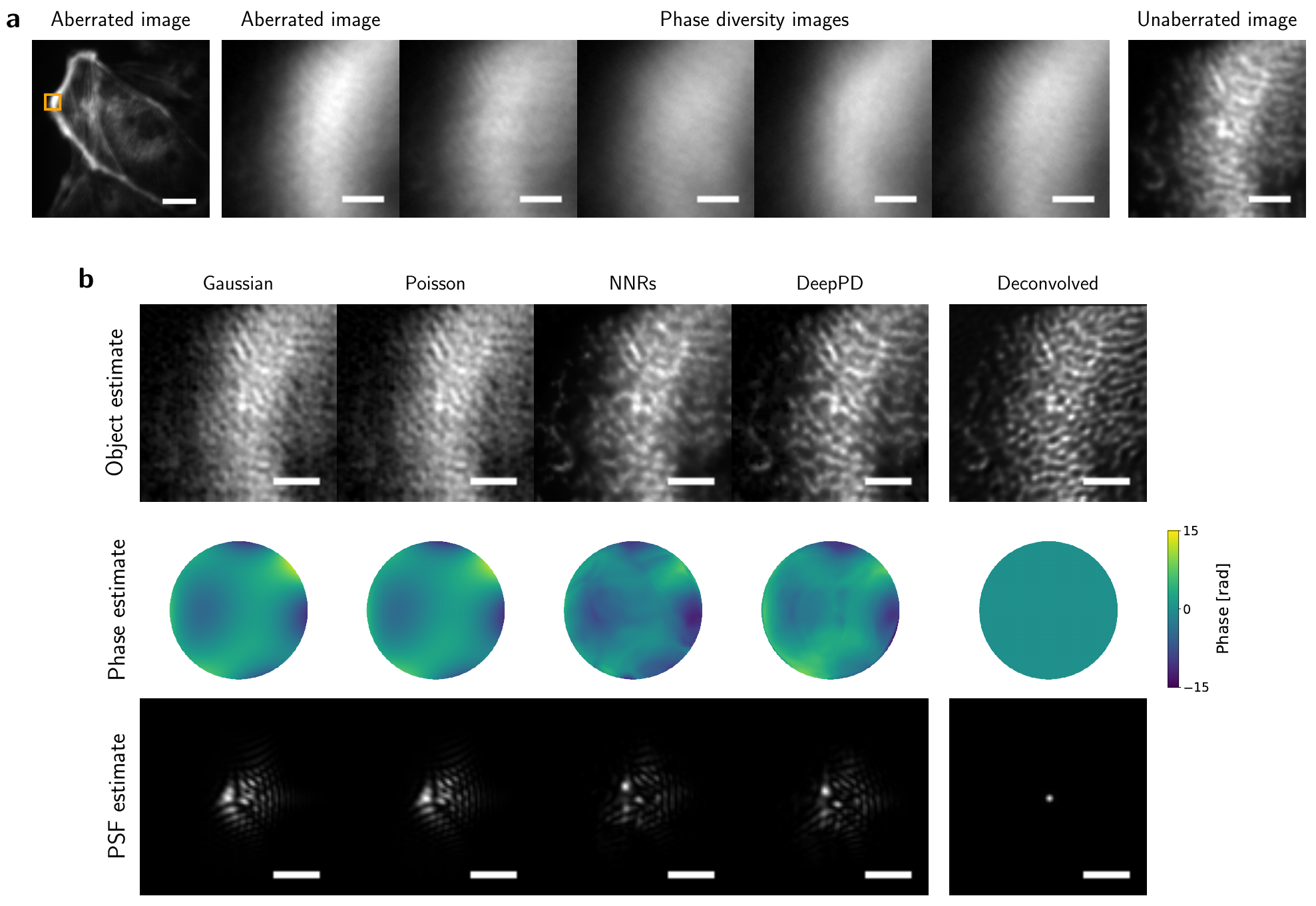}
    \caption{Performance comparison of various methods on a biological sample of myosin filaments (fixed PtK2 cell immunolabeled against myosin heavy chain).
    \panel{a} Recorded images. Left to right: Aberrated image, zoom-in of the aberrated image (region of interest indicated by the white box in the full image), four phase-diversity images, and the unaberrated image for comparison. Scale bars: $\SI{20}{\um}$ (first image), $\SI{2}{\um}$ (others).
    \panel{b} Object estimate (top row), phase estimate (middle row) and corresponding PSF (bottom row) obtained by various methods (left to right: Gauss--Newton algorithm, Poisson algorithm, neural representations assuming a linear mirror response (NNRs), and neural representations using the trained mirror model (DeepPD)). The last column shows an object estimate obtained by Richardson--Lucy deconvolution (20 iterations) of the unaberrated image, flat phase and corresponding PSF for comparison. Scale bars: $\SI{2}{\um}$.
    }
    \label{fig:myosin}
\end{figure}

\paragraph{Neural representations and a trained mirror network improve phase retrieval and object estimation on myosin samples.}
As the calibration samples from the Argo-HM slide provide only a limited set of structured patterns, we further investigated the performance of our approach on biological samples.
We first analyzed a previously published phase-diversity dataset of myosin filaments in fixed PtK2 cells immunolabeled against myosin heavy chain \cite{Johnson2024a}.
Fig.~\ref{fig:myosin}\subpanel{a} shows the recorded aberrated image, and a zoom-in (indicated by the orange box) of the aberrated image, the four phase-diversity images, and the unaberrated image for comparison.
Fig.~\ref{fig:myosin}\subpanel{b} shows the object estimate in the region of interest, the phase estimate, and the corresponding PSF obtained by the Gaussian, Poisson, NNRs, and DeepPD methods.
While the object estimates obtained by the Gaussian and Poisson methods are noisy and obstruct fine details of the sample structure, the two object estimates based on NNRs and DeepPD yield more accurate reconstructions, with DeepPD achieving the best reconstruction of high-frequency features.
For comparison, an object estimate based on Richardson--Lucy deconvolution of the unaberrated image is shown in the last column.
Interestingly, the object estimate obtained by our DeepPD method provides sharper features than the unaberrated image without deconvolution.
It is also evident that the PSFs obtained by the Gaussian and Poisson methods are similar to each other but distinctively different from the PSFs obtained by the NNRs and DeepPD methods.
This difference can be explained by the fact that both the Gaussian and Poisson method rely on a Zernike representation of the phase that is based on the first $21$ Zernike modes, while the NNRs and DeepPD methods are not limited to a specific basis set and can capture more complex phase structures.
The estimated PSFs from NNRs and DeepPD show similar features and only subtle differences.
While the Gaussian approach achieves a high discrete cosine transform (DCT) norm in real-time ($<$\SI{1}{s}) for a 512$\times$512 image region, DeepPD achieves superior performance at longer runtime ($\sim$\SI{20}{s}), while the Poisson method in its current implementation takes significantly longer ($>$\SI{100}{s}) to converge (Suppl.~Fig.~\ref{fig:speed-comparison}).

\begin{figure}
    \centering
    \includegraphics[width=\textwidth]{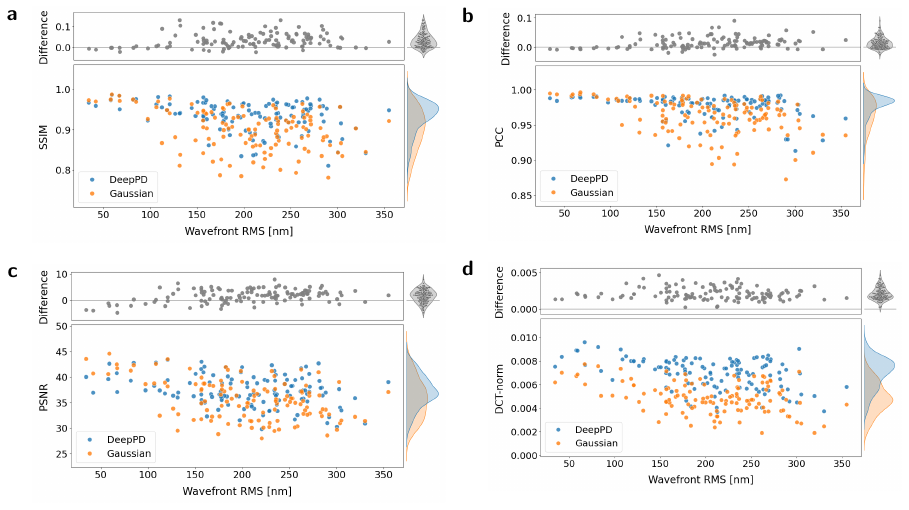}
    \caption{Image quality metrics for $121$ myosin datasets. \panel{a} Structural similarity index (SSIM). \panel{b} Pearson correlation coefficient (PCC). \panel{c} Peak signal-to-noise ratio (PSNR). \panel{d} Discrete cosine transform (DCT) norm.
    For SSIM and PCC, the maximum possible value is 1, which is indicated by the dashed horizontal lines.
    The lower subpanels shows the respective metrics for the Gaussian method (orange) and DeepPD (blue).
    The top subpanels show the differences between the two.
    The respective density distributions are shown on the right.
    For calculating SSIM, PCC and PSNR, the object estimate was convolved with the microscope's system PSF to obtain an estimated image; the reference image was the recorded unaberrated image.
    The x-axis shows the estimated wavefront RMS of the phase aberration for each dataset.
    }
    \label{fig:myosin-metrics}
\end{figure}

For a more quantitative comparison over multiple samples and different aberrations, we further recorded a set of $121$ datasets of immunolabeled myosin under the same experimental conditions with random phase aberrations introduced by the deformable mirror.
We calculated various image quality metrics for assessing the quality of the object estimates obtained by the Gaussian and DeepPD methods (Fig.~\ref{fig:myosin-metrics}), including the structural similarity index (SSIM), the Pearson correlation coefficient (PCC), the peak signal-to-noise ratio (PSNR), and the DCT norm.
For all metrics, higher values indicate better performance. The SSIM and PCC have a maximum value of 1, while PSNR and DCT norm have no upper limit.
In the subplots showing the differences in the metrics between the methods, positive values (i.e., values above the zero line) indicate better performance by our DeepPD method.
The SSIM, PCC and PSNR are all relative metrics and require a (ground truth) reference image for comparison.
As no ground truth for the object is available, we convolved the obtained object estimates with the microscope's PSF to obtain an estimated image, and used the recorded unaberrated image as the reference image.
The DCT norm is the only metric that does not rely on a reference image, and was calculated directly from the object estimates.
Notably, optimizing images for the DCT norm was reported to achieve the best results for other derived metrics (e.g., focus localization accuracy, signal-to-background ratio) in microscopy applications \cite{Royer2016}.
For RMS wavefront errors below $\SI{100}{\nm}$, the SSIM and PCC values for the myosin datasets are comparable between the Gaussian method and DeepPD. At higher RMS wavefront errors, our method outperforms the Gaussian approach.
PSNR values are slightly higher for the Gaussian method at low aberration magnitudes, but fall below the values obtained by our method for higher aberrations.
The DCT norm was calculated for frequency values within the microscope's numerical aperture, and is consistently higher for the object estimates obtained by our method compared to the Gaussian method, indicating a higher content in resolvable frequencies and less noise in the object estimates.

%% file: sections/03_discussion.tex
%%% Discussion

Optical aberrations---induced either by imperfections of the imaging system or the sample itself---remain a key limitation in light microscopy, as they prevent resolving sample features at the diffraction limit.
DeepPD addresses this problem by building on phase diversity, an indirect wavefront sensing technique that that leverages the additional information provided by the phase-diversity images and the knowledge of the associated diversity phases. However, if there is a mismatch between the assumed diversity phases and what is delivered by the deformable mirror, this is a source of error that hinders algorithm performance.

We introduced a neural network to model the response of a deformable mirror used for applying the phase diversities as part of DeepPD.
Accurate modeling of the mirror response allowed us to improve the phase-diversity estimates and, consequently, the estimates of the phase aberration and the object.
In future work the phase-to-voltage model can be used to apply the desired phase diversities, and further the phase aberration correction, more accurately. 
Depending on the used deformable mirror (especially its number of actuators), the desired phase, however, might not be perfectly achievable as the deformation space of the mirror is limited. In this case, it is still advisable to use the voltage-to-phase model to get a more accurate estimate of the induced phase.

Over time, the response of the mirror may slightly change due to changes in the environment such as temperature or humidity, or due to mechanical wear. To ensure high-quality calibration of the mirror, the mirror model can be fine-tuned with additional training data.

Using the phase-diversity approach, we reliably retrieved the phase aberration for RMS wavefront error values up to around $\SI{350}{\nm}$ (Fig.~\ref{fig:myosin-metrics}), which is higher than previously reported limits of around \SI{200}{\nm} to \SI{250}{\nm} \cite{Johnson2024a}.
To improve the estimates of the phase aberrations induced by the mirror for the training data, we initialized the object and phase estimates with the unaberrated image and the expected phase from the linear mirror model or previous iterations of the trained mirror model.
This information---which is available for the training data, but not in the general case where aberrations are induced by the sample itself---helps the optimization algorithm to converge to the correct solution even for higher aberration magnitudes.
Still, at some point the signal-to-noise ratio of the images will be too low to reliably estimate the phase aberration, preventing the acquisition of reliable training data beyond a certain wavefront error.
Therefore, the mirror model used for calibration will be most accurate for phase aberrations in a range that can be reliably estimated based on the phase-diversity images.
For larger aberrations, the model would either be misled by unreliable phase estimates if training data at those ranges is included, or would need to generalize beyond the training data, which may lead to inaccuracies in the predictions. Therefore, the calibration based on the mirror model should only be used for aberrations within the range of reasonable training data.

Our approach requires a set of only four additional phase-diversity images (as in \cite{Johnson2024a}), which is significantly less than the roughly $100$ images required for reliable phase retrieval based on neural representations with random phase diversities as reported in NeuWS \cite{Feng2023}. This allows rapid imaging of moving samples and minimal light exposure, which is especially important for live-cell imaging applications.
Ideally, the retrieved phase aberration is corrected directly at acquisition time via the deformable mirror, yielding a sharp recorded image with high signal-to-noise ratio.

With the currently achieved reconstruction times of our method of a few seconds for medium-sized images of $256\times 256$ or $512\times 512$, real-time correction of the phase aberration is feasible for samples moving at slow speeds.
For rapidly moving samples, an online correction may not always be feasible. In these cases, we showed that DeepPD can nonetheless achieve high-quality object estimates in post-processing, providing valuable information about the sample that could be impossible to retrieve from the aberrated images alone.
Future research on optimization of the computational algorithm, and hardware improvements will further enhance the analysis speed, promising real-time correction of the phase aberration even in rapidly changing samples.

In summary, DeepPD combines neural representations, selected phase diversities, and a learned mirror model to jointly estimate object structure and phase aberrations from only five images, achieving accurate reconstructions across a wide range of aberrations beyond the limits of previous approaches.

%% file: sections/04_methods.tex
\subsection{Sample preparation and microscope setup}
The imaged sample for Fig.~\ref{fig:argolight} was an Argo-HM (High Magnification) V2 slide (Argolight) which contains a special glass substrate with fluorescent patterns embedded inside. We imaged the letter G of the pattern featuring the word ARGOLIGHT.
All cell data analyzed was recorded previously and is partially published in \cite{Johnson2024a}. In short, PtK2 cells were fixed and immunolabeled against myosin heavy chain. Images were recorded on a widefield adaptive optics system with a DM (Imagine Optic, MIRAO 52ES) and a SH sensor (Imagine Optic, HASO4 First Shack–Hartmann wavefront sensor, IO-6WFS201) in the emission path. The pixel-size of the EM-CCD camera (Oxford Instruments, iXon Ultra 888, DU-888U3-CS0-\#BV) in image space was $\pixelsize=\SI{104}{\nm}$, the central emission wavelength $\lambda=\SI{532}{\nm}$, and the numerical aperture of the objective was $\NA=1.2$.
All details of sample preparation, microscope setup and imaging parameters can be found in \cite{Johnson2024a}.

\subsection{Phase-diversity analysis}
We compared the performance of phase-diversity analysis using several different algorithms described below, which are based on either analytic derivatives or automatic differentiation.
The deconvolved images for comparison in \ref{fig:argolight}\subpanel{b} and \ref{fig:myosin}\subpanel{b} were calculated using the scikit-image \cite{Walt2014} implementation of Richardson--Lucy deconvolution with $20$ iterations.

\subsubsection{Gauss--Newton algorithm}
The Gauss--Newton algorithm for phase retrieval assumes a Gaussian noise distribution in the data and relies on analytic derivatives of the theoretical point spread function. Details are described in \cite{Vogel1998, Johnson2024a}.
In short, the unknown phase aberration is represented by a linear combination of 18 Zernike polynomials (i.e., the first 21 Zernike polynomials excluding piston, tip and tilt). The Zernike coefficient estimate $\mathbf{c}$ is iteratively updated by
\begin{equation}
    \mathbf{c}_{n+1} = \mathbf{c}_{n} + \Delta \mathbf{c}, \quad \textrm{where }\Delta \mathbf{c} = -\mathbf{H}^{-1} \mathbf{g},
\end{equation}
with $\mathbf{H}$ and $\mathbf{g}$ being the Hessian matrix and gradient of the cost function $J$, respectively. The cost function is defined as
\begin{equation}
    J = \sum_{k=1}^{N} |D_k|^2 - \frac{\sum_{k=1}^{N} |D_k^* S_k(\phi)|^2}{\gamma + \sum_{k=1}^{N}|S_k(\phi)|^2},
\end{equation}
where $D_k$ are the recorded images in frequency space, $S_k(\phi)$ the optical transfer functions, and $\gamma$ a regularization parameter set to $10^{-4}$. For the calculation of the gradient and Hessian matrix see equations (A4) and (A6) in reference \cite{Johnson2024a}, respectively.
The object estimate was retrieved as
\begin{equation}
    \objest = \fourierinv{\frac{\sum_{k=1}^{N}S^*_k(\phaseest) D_k}{\gamma + \sum_{k=1}^{N} |S_k(\phaseest)|^2}}{\bigg} \, ,
\end{equation}
where $\phaseest$ is the phase aberration estimate.

\subsubsection{Poisson algorithm}
The Poisson algorithm assumes a Poisson noise distribution in the data. The details of the algorithm are described in \cite{Reiser2023}. In short, the estimates of the Zernike aberration coefficients of the unknown phase aberration and the object estimate are alternately updated in each iteration following equations (13) and (15) in reference \cite{Reiser2023}.
The number of Zernike modes for the analysis was set to 18 (i.e., the first 21 excluding piston, tip and tilt).
The estimated Zernike coefficients were initialized with random values from a normal distribution with a mean of $0$ and a standard deviation of $10^{-4}$.
The object estimate was initialized with a constant value of $1$.
For the line search, the step size was initialized with $3\times10^4$ and reduced by a factor of $0.3$ in each iteration.
The maximum number of iterations for each line search was set to $10$.
The algorithm was run for 700 iterations at which point no further improvement in the resulting estimates was observed.

\subsubsection{Neural representations}
The phase-diversity analysis using neural representations is adapted from reference \cite{Feng2023}. As in NeuWS \cite{Feng2023}, the object and phase estimate are represented by a neural network.
We refer to the full approach using neural representations and the trained mirror model as DeepPD, and distinguish it from a variant using a linear mirror model, referred to as NNRs.

\paragraph{Object estimate} For the object estimate, we use a $K$-planes representation \cite{FridovichKeil2023}, with a learnable $N\times N$ feature map of dimension 32. This feature map is interpolated at the query coordinate positions and the obtained features are the input to a multi-linear perceptron (MLP) with ReLU activation functions, 2 hidden layers and a hidden dimension of 16. The output of the MLP is the object estimate value at the given coordinate. We evaluate the MLP at coordinate positions of a $N\times N$ grid to obtain the object estimate over the whole region of interest. As a modification to the NeuWS network, the output of the last layer of the MLP is squared in order to prevent negative object values, which facilitates training and leads to improved object estimates.

\paragraph{Phase estimate} The phase estimate is represented by a neural network with a spatial frequency coordinate grid as input. We evaluate 18 Zernike polynomials at these grid positions (i.e., the first 21 Zernike polynomials excluding piston, tip and tilt). Further, the values of this Zernike basis are passed into a MLP with LeakyReLU activation functions, 2 hidden layers and a hidden dimension of 32. The MLP parameters are initialized with He Normal initialization \cite{He2015}.

\paragraph{Representation of phase diversities}
In all microscopy experiments, the phase diversities were applied via the DM. In our reconstruction algorithm, we use two different ways to account for the phase diversities:
(i) the phase of each diversity is calculated as a linear combination of the influence functions,
(ii) in DeepPD, the phase is predicted from the applied voltages using the trained mirror model.

\subsubsection{Simulation of image formation}
The microscope image is formed by convolution of the object with the point spread function. The estimated diversity images in each iteration are hence given by
\begin{equation}
    \widehat{I}_k = \objest * | \mathcal{F}^{-1} (P\, e^{i(\phaseest+\phasediv_k)} ) | ^2
    =: \objest * \PSF(\phaseest+\phasediv_k),
    \label{eq:image-formation}
\end{equation}
where $\objest$ is the current object estimate, $P$ the pupil function, $\phaseest$ the phase estimate, $\phasediv_k$ the $k$th phase diversity, and $\PSF(\phaseest + \phasediv_k)$ the point spread function (PSF) with the phase aberration $\phaseest + \phasediv_k$.
Note that in our case $\phasediv_0\equiv0$ as the first image is the aberrated image without additional phase diversity applied to the unknown phase aberration.
In practice, we evaluate Eq.~\eqref{eq:image-formation} in Fourier space via the optical transfer function:
\begin{equation}
    \fourier{\widehat{I}_k} = \fourier{\objest} \cdot \fourier{ \PSF(\phaseest+\phasediv_k) } =: \fourier{\objest} \cdot \OTF(\phaseest+\phasediv_k).
\end{equation}

\subsubsection{Loss}
The loss function for the phase-diversity analysis is defined as the mean squared error (MSE) between the recorded phase-diversity images and the simulated images based on the object and phase estimates:
\begin{equation}
    \loss{PD} = \frac{1}{N} \sum_{k=1}^{N} \big\| I_k - \widehat{I}_k \big\|^2,
\end{equation}
where $N$ is the number of phase-diversity images ($N=5$ in our case), $I_k$ the recorded $k$th diversity image, and $\widehat{I}_k$ the corresponding simulated image based on the object and phase estimates.
Moreover, we penalize estimated phase values that are above $\num{200}\pi~\si{\radian}$ to improve the convergence of the optimization.
Phase aberration values exceeding the threshold are penalized using the L2 norm of their deviations beyond the threshold.
Note that a value of $\num{200}\pi~\si{\radian}$ corresponds to a phase shift of $\SI{53.2}{\um}$ (assuming a wavelength of $\SI{532}{\nm}$), which is well above the range of aberrations that could be reliably retrieved by the phase-diversity analysis.
The hyperparameters for weighting $\loss{PD}$ and the phase threshold were set to $10^7$ and $1$, respectively.
Positive values of the object estimate are naturally ensured by the neural representation of the object, as the output of the last layer of the MLP is squared.

\subsubsection{Training}
The neural representation of the object was initialized over $100$ iterations with the aberrated image based on an MSE loss using the Adam optimizer with a learning rate of $10^{-2}$.
We further trained the object and phase neural representations over 700 iterations, using the Adam optimizer with a learning rate of $10^{-2}$ for both neural representations.

\subsection{Neural network for deformable mirror}

\subsubsection{Training data}
The training data for the mirror model needs to be a pair of input voltages and induced phase aberrations. As a ground truth for the phase aberration does not exist, we use phase aberration estimates retrieved by our phase-diversity analysis using the neural representations as a proxy.

For this, we recorded a set of $\num{3930}$ phase-diversity image stacks---each consisting of one aberrated image and four associated astigmatism phase-diversity images---on $\SI{100}{\nm}$ bead samples with random phase aberrations applied by the DM.
The random aberrations were sampled from the first 21 Zernike polynomials excluding piston, tip, tilt and defocus modes.
For each aberration, each Zernike coefficient was set to $0$ with a probability of $\SI{70}{\percent}$; otherwise, a random coefficient value was drawn from a uniform distribution in the interval $[-A, A]$, where $A$ is a given maximum amplitude which was varied between $0.1$ and $\SI{0.8}{\um}$.
The obtained Zernike coefficients were converted to input voltages using the linear mirror model.

As our aim is to get the closest possible estimate of the phase aberration, we use all available information we have for retrieving the phase estimate for the mirror training data. Hence, we initialize the object and phase estimates with the best initial guess we have. Note that this cannot be done for phase retrieval in general, as the object and phase are unknown and to be retrieved.
The object estimate is initialized with the unaberrated image (i.e., the recorded image without inducing an additional aberration with the DM).
The phase aberration estimate is initialized with the expected phase as determined from the mirror influence functions (assuming a linear mirror response). Both the object and phase estimate initialization were done over 100 epochs each using an Adam optimizer with a learning rate of $10^{-2}$.

\subsubsection{Architecture}\label{sec:methods-nn-architecture}
The mirror model consists of two parts, one network predicting the phase aberration induced by a given set of voltages (\textit{voltage-to-phase network}), and a second network predicting the voltages required to induce a desired phase aberration (\textit{phase-to-voltage network}).
A sketch of the architecture of the mirror model is depicted in Fig.~\ref{fig:mirror-model}.
For the voltage-to-phase network, the input is a 52-element vector representing the voltages applied to each of the 52 actuators of the DM.
This input is passed through 2 fully connected layers: the first layer transforms the input into a 64-dimensional feature vector with a ReLU activation, and the second expands this to a $\num{4096}$-dimensional vector ($64\times 64$) followed by ReLU activation.
The result is reshaped to a size of $64\times 8 \times 8$, corresponding to 64 channels and $8\times 8$ spatial dimension to prepare the input for the convolutional layers.
The following three convolutional blocks each consist of an upsampling layer (bilinear interpolation) to increase spatial resolution and three convolutional layers with ReLU activations.
The blocks progressively reduce the number of channels from 64 to 16, to 4, and down to 2, while increasing the spatial dimensions by a factor of 4 in the first two blocks and 2 in the last block.
Finally, a $1\times 1$ convolutional layer reduces the channel dimension to the desired output size of a single channel representing the phase aberration.
The output is element-wise multiplied by a pupil mask to account for the microscope aperture.

The phase-to-voltage network is the reverse of the voltage-to-phase network.
The input is a single channel image of the desired phase aberration, which is first masked with the pupil aperture and passed through a $1\times 1$ convolutional layer that increases the number of channels from 1 to 2.
Subsequently, this is passed through three convolutional blocks, each consisting of three convolutional layers with ReLU activations and a downsampling layer using average pooling, reducing the spatial dimensions by a factor of 2 in the first and by 4 in each of the following blocks.
The number of channels increases from 2 to 4, then to 16, and finally to 64.
The output of the last convolutional block is reshaped to size $64\times 64$ and passed through two fully connected layers (with ReLU activation after the first layer)---the first reduces it to a 64-dimensional vector, and the second predicts the 52-element vector of voltages.

\subsubsection{Loss}
Both the voltage-to-phase and phase-to-voltage networks are trained simultaneously.
For training, we used the Adam optimizer with a learning rate of $10^{-4}$ and a batch size of $8$.
The total loss $\Lmirror$ for training of the mirror model is comprised of several components and calculated as
\begin{equation}
    \Lmirror = \Lsupervision + \Lcycle + \Lregularization,
\end{equation}
consisting of the supervision loss $\Lsupervision$, the cycle-consistency loss $\Lcycle$, and a regularization term $\Lregularization$.
The loss components are defined as follows:

\begin{description}
    \item[Supervision loss:] The supervision loss $\Lsupervision$ consists of two components: (i) the sum of the MSE between the phase estimate obtained from the phase-diversity analysis of the training data, and the predicted phase aberration by the mirror model, and (ii) the MSE between the input voltages and the predicted voltages to require the desired phase aberration.
    The hyperparameters for weighing the phase and voltage prediction were set to $1$ and $0.01$, respectively.
    \item[Cycle-consistency loss:] When estimating the phase aberration from the input voltages, and subsequently predicting the voltages required to induce that phase, the final voltages should correspond to the initial input voltages. Vice versa, when predicting the voltages from a desired phase aberration, and then estimating the phase from these voltages, the final phase should correspond to the desired phase. The cycle-consistency loss $\Lcycle$ enforces these constraints by calculating the MSE between the input voltages and the predicted voltages, and between the desired phase and the predicted phase.
    The hyperparameters for weighing the phase and voltage cycle-consistency were set to $10$ and $0.1$, respectively.
    \item[Regularization:] The regularization loss $\Lregularization$ penalizes phase values that are above the maximum range of the DM. For training, we set the phase value threshold to $\num{200}\pi~\si{\radian}$.
    The regularization loss was weighted with a hyperparameter of $1$.
\end{description}

\subsubsection{Iterative improvement of the mirror model}
As the initial phase retrieval for the training data assumes a linear mirror model, the phase estimate is affected by residual errors.
Therefore, we decided on an iterative scheme, where we first train the mirror model with the initial phase estimates, then reanalyze the training dataset with phase diversities predicted from the trained mirror model, and subsequently retrain the mirror model with the newly obtained phase estimates.
While the underlying experimental dataset for the training data remained the same, the estimate of the phase aberration induced by the given voltage was updated in each cycle.
This process was repeated for three cycles, after which the difference between the phase prediction from the mirror model and the prediction from re-analyzing the training data converged.
The mirror model was trained or fine-tuned for $400$ epochs in each cycle.

\subsection{Shack--Hartmann phase detection}
For proof of concept, we validated our approach against the gold standard method in phase retrieval, i.e., a SH wavefront sensor, which consists of an array of microlenses measuring the phase aberration.
The SH sensor was installed in a separate light path of the microscope.
For all experiments, the exposure time was set to $\SI{1}{s}$.
The Imagine Optic Wavekit software was used to retrieve the zonal and modal representation of the phase aberration from the raw data.
For details of the SH sensor, microscope setup, and sensor control see \cite{Johnson2024a}.

\subsection{Metrics for performance comparison}
The various metrics used for performance comparison of the different phase-diversity analysis methods are described below.

\paragraph{Root Mean Square Wavefront Distortion}
The root mean square wavefront distortion $\RMSWD$ is calculated as the root mean square error with respect to a flat wavefront evaluated at $n$ positions of a discretized pupil according to
\begin{equation}
    \RMSWD = \sqrt{\frac{1}{N} \sum_{n=1}^{N} | \phaseest_{(n)} |^2},
\end{equation}
where $\phaseest_{(n)}$ is the phase estimate at the $n$th position of the pupil.

\paragraph{SSIM}
The structural similarity index measure (SSIM) compares the similarity between two images based on luminance, contrast, and structure. The SSIM index is calculated as
\begin{equation}
    \SSIM = \frac{(2\mu_1\mu_2 + c_1)(2\sigma_{1,2} + c_2)}{(\mu_1^2 + \mu_2^2 + c_1)(\sigma_1^2 + \sigma_2^2 + c_2)},
\end{equation}
where $\mu_1$ and $\mu_2$ are the mean values of the two images, $\sigma_1$ and $\sigma_2$ the standard deviations of the image values, $\sigma_{1,2}$ the covariance, and $c_1=(0.01 D)^2$ and $c_2=(0.03 D)^2$ are stabilization constants with $D$ being the dynamic range of the pixel values ($D=1$ for normalized images). The range of the SSIM index is $[-1, 1]$, with $1$ indicating perfect similarity, $0$ indicating no similarity, and $-1$ indicating perfect anti-correlation.
SSIM values were calculated using the scikit-image \cite{Walt2014} implementation.

\paragraph{PSNR}
The peak signal-to-noise ratio (PSNR) compares the quality of an image to a reference image and is calculated as
\begin{equation}
    \textrm{PSNR} = 10 \log_{10} \left( \frac{D^2}{\MSE} \right),
\end{equation}
where $D$ is the range of the pixel values of the reference image and $\MSE$ the mean squared error between the two images.
PSNR values were calculated using the scikit-image \cite{Walt2014} implementation.

\paragraph{DCT-norm}
The discrete cosine transform (DCT) norm is an image quality metric that quantifies high-frequency content. To avoid noise contributing to the quality metric, we calculate the DCT-norm on spatial frequencies within the resolvable range of the microscope. We calculated the DCT-norm of image $I$ as described in \cite{Royer2016}:
\begin{equation}
    \DCT_{r_0}(I) = -\frac{2}{r_0^2} \sum_{x+y<r_0} \Bigg(\frac{|\dct (I)_{x,y}|}{\Vert \dct (I)\Vert} \; \abslog{2} \bigg( \frac{\dct (I)_{x,y}}{\Vert \dct (I) \Vert} \bigg) \Bigg), 
\end{equation}
with $r_0=2\NA/\lambda$ being the diffraction limit of the microscope, $\dct (I)$ the discrete cosine transform of the image, $\Vert.\Vert$ the $L_2$ norm, and $\abslog{2}$ is defined as $\log_2(|x|)$ for $x\neq 0$, and $0$ otherwise.

\paragraph{PCC}
The Pearson correlation coefficient (PCC) measures the linear correlation between the values of two images and is calculated as
\begin{equation}
    \textrm{PCC} = \frac{\sum_{i=1}^{N} (I_{1,i} - \mu_1)(I_{2,i} - \mu_2)}{\sqrt{\sum_{i=1}^{N} (I_{1,i} - \mu_1)^2} \sqrt{\sum_{i=1}^{N} (I_{2,i} - \mu_2)^2}},
\end{equation}
where $I_{1,i}$ and $I_{2,i}$ are the $i$th pixel values of the first and second image, respectively, and $\mu_1$ and $\mu_2$ are the mean values of the two images.
The range of the PCC is $[-1, 1]$, with $1$ indicating perfect positive correlation, $0$ indicating no correlation, and $-1$ indicating perfect negative correlation.

\subsection{Computational hardware}
All computations and model training were run on a Ubuntu-based workstation featuring an Intel Xeon w9-3495X 56-core CPU, \SI{512}{GB} system RAM, and an NVIDIA RTX A6000 with \SI{48}{GB} RAM.

%% file: sections/SI.tex
\section*{Supplementary Information}

\vspace{30pt}

\begin{figure}[!htb]
    \centering
    \includegraphics[width=0.6\textwidth]{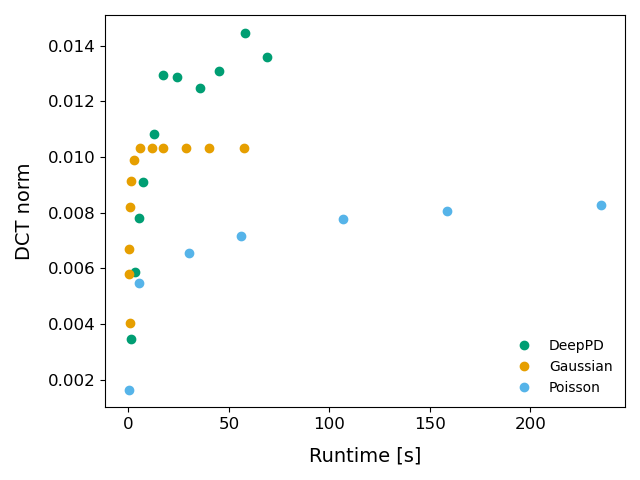}
    \vspace{-5pt}
    \caption{Comparison of runtimes and corresponding achieved DCT norm of the object estimate for the Gaussian, Poisson, and DeepPD methods for a 512$\times$512 image region of the myosin dataset shown in Fig.~\ref{fig:myosin} in the main text.
    }
    \label{fig:speed-comparison}
\end{figure}